\documentclass[12p]{article}
\usepackage{latexsym}
\usepackage{color}
\usepackage{amsmath}
\usepackage{epsfig}
\usepackage{hyperref}
 \usepackage{dcolumn}
   \usepackage{threeparttable}

\newcommand{\ortala}[1]{\begin{center}#1\end{center}}

\newcommand{\sandd}[1]{\left\langle #1\right\rangle}

\newcommand{\integ}[3]{{{\underset{#1 }{\overset{#2}{\displaystyle\int}}}#3}}

\newcommand{\summ}[3]{{{\underset{#1 }{\overset{#2}{\displaystyle\sum}}}#3}}

\newcommand{\re}[1]{(\ref{#1})}

\newcommand{\eq}[2]{\begin{equation}\label{#1}  #2\end{equation}}

\newcommand{\paran}[1]{\left(#1\right)}

\newcommand{\sch}[1]{Schrodinger}

\newcommand{\komb}[2]{\paran{\begin{array}{c} #1 \\ #2 \end{array}}}

\linethickness{0.6pt}

\begin{document}

\ortala{\textbf{Critical Behavior of the 3D anisotropic quantum Heisenberg model in a trimodal random field distribution}}

\ortala{\textbf{\"Umit Ak\i nc\i \footnote{umit.akinci@deu.edu.tr}}}

\ortala{\textit{Department of Physics, Dokuz Eyl\"ul University,
TR-35160 Izmir, Turkey}}

\section{Abstract}

Effect of the trimodal random magnetic field distribution on the phase
diagrams of the anisotropic quantum Heisenberg model has been
investigated for three dimensional lattices with effective field theory (EFT) for a two spin cluster.
Variation of the phase diagrams with the random magnetic field
distribution parameters has been obtained and the effect of the anisotropy in the exchange interaction on the phase diagrams has been investigated in detail.
Particular attention has been devoted on the behavior of the tricritical points with random magnetic field distribution.
Keywords: \textbf{Quantum anisotropic Heisenberg model; random
magnetic field; trimodal distribution}

\section{Introduction}\label{introduction}

Recently there has been growing theoretical interest in the random field lattice spin models. For instance,
Ising model in a quenched random field (RFIM) has
been studied over three decades. The model was introduced for the first time by Larkin
\cite{ref1} for superconductors and later generalized by Imry and Ma \cite{ref2}. Diluted antiferromagnets (such as $Fe_xZn_{1-x}F_2$, $Rb_2Co_xMg_{1-x}F_4$ and $Co_xZn_{1-x}F_2$)
in a homogenous magnetic field behave like ferromagnetic systems in the presence of random fields \cite{ref3,ref4}.
Beside this, a rich class
of experimentally accessible disordered systems can be described by RFIM, such
as structural phase transitions in random alloys, commensurate charge- density-wave systems with impurity pinning, binary fluid mixtures in random porous
media, and the melting of intercalates in layered compounds, such as $TiS_2$\cite{ref5}.
RFIM generally mimics the phase transitions and interfaces in random media
\cite{ref6,ref7}, e.g prewetting transition on a disordered substrate can be mapped onto a
2D RFIM problem \cite{ref8}. Also, RFIM has been applied in order to describe critical
surface behavior of amorphous semi-infinite systems \cite{ref9,ref10}.

Random field distribution of the magnetic field produces drastic effects on the phase
diagrams and related magnetic properties of the system. It has been shown that Ising systems under
the influence of discrete symmetric distributions,
like bimodal \cite{ref11} and trimodal \cite{ref12} distributions, show tricritical behavior,
while  continuous symmetric distributions like Gaussian distribution \cite{ref13} exhibit only second order transitions.

On the other hand, as far as we know, there have been less attention
paid on the random field effects on the Heisenberg model, which is more realistic model than the Ising model for the spin systems.
Albuquerque and Arruda \cite{ref14} studied the effect of the bimodal random field distribution on the phase transition characteristics of the spin-1/2 isotropic classical  Heisenberg model and they found tricritical behavior within the EFT formulation for the two spin cluster which is abbreviated as EFT-2. Oubelkacem et al., studied the same system with another approach, namely EFT with probability distribution technique and they obtained similar results \cite{ref15}.
Albuquerque et al. \cite{ref16} treated the same system with amorphisation effect, again with the EFT-2 formulation.  Recently,
Sousa et al. have studied the effect of the bimodal random field distribution on phase transition characteristics of the isotropic -classical and quantum- spin-1/2 Heisenberg model within the EFT-2 formulation and also they found a tricritical behavior \cite{ref17}. All these works have been restricted to the spin-1/2 isotropic Heisenberg model with bimodal random field distribution and they concluded that tricritical behavior exists in this system as in Ising model with bimodal random field distribution. They utilized an EFT  which is characterized  by differential operator technique introduced by Honmura and Kaneyoshi for Ising systems \cite{ref18}. EFT approximation can provide results that are superior to those obtained
within the traditional mean field approximation, due to the consideration of self spin correlations which are omitted in the mean field approximation. EFT for a typical Ising system starts by constructing a finite cluster of spins which represents the system. Callen-Suzuki spin identities \cite{ref19,ref20} are the starting point of the EFT for the one spin clusters. If one expands these identities with differential operator technique, multi spin correlations appear, and in order to avoid from the mathematical difficulties, these multi spin correlations are often neglected by using decoupling approximation \cite{ref21}. Working with larger finite clusters will give more accurate results. Callen-Suzuki identities have been generalized to two spin clusters in Ref. \cite{ref22} (EFT-2 formulation). This EFT-2 formulation has been successfully applied to a variety of systems, such as quantum spin-1/2 Heisenberg ferromagnet \cite{ref23,ref24}  and antiferromagnet \cite{ref25} systems,  classical n-vector model \cite{ref26,ref27}, and spin-1 Heisenberg ferromagnet \cite{ref28,ref29}.

The aim of this work is to investigate the effect of the symmetric  discrete random field distributions (bimodal and trimodal) on the phase transition characteristics of a spin-1/2 anisotropic quantum Heisenberg model on simple cubic and body centered cubic lattices. Quantum Heisenberg model can take into account the quantum fluctuations which dominates the thermal fluctuations in the low temperatures. Thus it is expected that it gives more reasonable results than the classical one at this low temperature region. We follow the EFT-2 formulation which is derived in Ref. \cite{ref23} for this system.

The paper is organized as follows: In Sec. \ref{formulation}, we
briefly present the model and  formulation. The results and
discussions are presented in Sec. \ref{results}, and finally Sec.
\ref{conclusion} contains our conclusions.

\section{Model and Formulation}\label{formulation}

We consider a lattice which consists of $N$ identical spins (spin-$1/2$) such that each of the spins has $z$ nearest neighbors. The Hamiltonian of the system is given by
\eq{denk1}{\mathcal{H}=-\summ{<i,j>}{}{\paran{J_x s_i^xs_j^x+J_y s_i^ys_j^y+J_z s_i^zs_j^z}}-\summ{i}{}{H_is_i^z}}
where $s_i^x,s_i^y$ and  $s_i^z$ denote the Pauli spin operators at a site $i$. $J_x,J_y$ and $J_z$ stand for the anisotropy in the exchange interactions between the nearest neighbor spins and $H_i$ is the longitudinal magnetic field at a site $i$. The first sum is carried over the nearest neighbors of the lattice, while the second one is over all the lattice sites. Magnetic field
is distributed on the lattice sites according to a trimodal distribution
function which is given by
\eq{denk2}{P\paran{H_i}=p\delta\paran{H_i}+\frac{1-p}{2}\left[\delta\paran{H_i-H_0}+\delta\paran{H_i+H_0}\right]
} where $p$ is a real number which provides $0\le p \le 1$, and $\delta$ stands for the delta function. The distribution given by Eq. \re{denk2}
covers a bimodal distribution for $p=0$ and reduces to the system with zero magnetic field (pure system) for $p=1$. According to the distribution given in Eq. \re{denk2}, $p$ percentage of the lattice sites are subjected to a magnetic field $H_i=0$, while half of the remaining sites are under the influence of a field $H_i=H_0$  whereas the field $H_i=-H_0$ acts on the remaining sites.

We use the two spin cluster approximation as an EFT formulation namely EFT-2 formulation\cite{ref23}. In this approximation, we choose two spins (namely $s_1$ and $s_2$) and treat interactions exactly in this two spin cluster. In order to avoid some mathematical difficulties we replace the perimeter spins of the two spin cluster by Ising spins (axial approximation) \cite{ref24}. After all, by using the differential operator technique and decoupling approximation (DA) \cite{ref21},
we get an expression for the magnetization per spin as
\eq{denk3}{
m=\sandd{\frac{1}{2}\paran{s_1^z+s_2^z}}=\sandd{\left[A_{x}+m B_{x}\right]^{z_0}
\left[A_{y}+m B_{y}\right]^{z_0}
\left[A_{xy}+m B_{xy}\right]^{z_1}} F\paran{x,y,H_0}|_{x=0,y=0}
} where each of $s_1$ and $s_2$ has number of $z_0$ distinct nearest neighbors and both of them have $z_1$ common nearest neighbors.
The coefficients are defined by
\eq{denk4}{
\begin{array}{lcl}
A_{x}=\cosh{\paran{J_z\nabla_x}}&\quad&
B_{x}=\sinh{\paran{J_z\nabla_x}}\\
A_{y}=\cosh{\paran{J_z\nabla_y}}
&\quad&
B_{y}=\sinh{\paran{J_z\nabla_y}}\\
A_{xy}=\cosh{\left[J_z\paran{\nabla_x+\nabla_y}\right]}&\quad&
B_{xy}=\sinh{\left[J_z\paran{\nabla_x+\nabla_y}\right]}\\
\end{array}
}
where $\nabla_x=\partial/\partial x$ and $\nabla_y=\partial/\partial y$ are the usual differential operators in the
differential operator technique. Differential operators act on an arbitrary function via
\eq{denk5}{\exp{\paran{a\nabla_x+b\nabla_y}}G\paran{x,y}=G\paran{x+a,y+b}}
with any constant  $a$ and $b$. The function in Eq. \re{denk3} is given by
\eq{denk6}{F\paran{x,y,H_0}=\integ{}{}{}dH_1dH_2P\paran{H_1}P\paran{H_2}f\paran{x,y,H_1,H_2}}
where
\eq{denk7}{f\paran{x,y,H_1,H_2}=\frac{x+y+H_1+H_2}{X_0}\frac{\sinh{\paran{\beta X_0}}}{\cosh{\paran{\beta X_0}}+\exp{\paran{-2\beta J_z}}\cosh{\paran{\beta Y_0}}}} and
\eq{denk8}{
X_0=\left[\paran{J_x-J_y}^2+(x+y+H_1+H_2)^2\right]^{1/2}, \quad Y_0=\left[\paran{J_x+J_y}^2+(x-y+H_1-H_2)^2\right]^{1/2}.
}
Here $\beta=1/(k_B T)$ where $k_B$ is Boltzmann
constant and $T$ is the temperature.

With the help of the Binomial expansion, Eq. \re{denk3} can be written as
\eq{denk9}{
m=\summ{p=0}{z_0}{}\summ{q=0}{z_0}{}\summ{r=0}{z_1}{}C^\prime_{pqr}m^{p+q+r}
} where the coefficients are
\eq{denk10}{
C^\prime_{pqr}=\komb{z_0}{p}\komb{z_0}{q}\komb{z_1}{r}A_x^{z_0-p}A_y^{z_0-q}A_{xy}^{z_1-r}B_x^{p}B_y^{q}B_{xy}^{r}F\paran{x,y,H_0}|_{x=0,y=0}
} and these coefficients can be calculated by using the definitions given in Eqs. \re{denk4} and \re{denk5}. Let us write Eq. \re{denk9} in more familiar form as
\eq{denk11}{
m=\summ{k=0}{z}{}C_{k}m^{k}
}
\eq{denk12}{
C_{k}=\summ{p=0}{z_0}{}\summ{q=0}{z_0}{}\summ{r=0}{z_1}{}\delta_{p+q+r,k}C^\prime_{pqr}
} where $\delta_{i,j}$ is the Kronecker delta. It can be shown from the symmetry properties of the function defined in Eq. \re{denk6} and operators defined by Eq. \re{denk4} that for even $k$, the coefficient $C_k$ is equal to zero.

For a given set of Hamiltonian parameters ($J_x,J_y,J_z$), temperature and field distribution parameters ($p,H_0$), we can determine the coefficients  from Eq. \re{denk12} and we can obtain
a non linear equation from Eq. \re{denk11}. By solving this equation, we can get the magnetization ($m$) for a given set of parameters and temperature. Since the magnetization is close to zero in the vicinity of the critical point, we can obtain a linear equation by linearizing the equation given in  Eq. \re{denk11} which allows us to determine the critical temperature. Since we have not calculated the free energy in this approximation, we can locate only second order transitions from the condition given as
\eq{denk13}{
C_1=1, \quad C_3<0
}The tricritical point at which second and first order transition lines meet can be determined from the condition
\eq{denk14}{
C_1=1, \quad C_3=0.
}

\section{Results and Discussion}\label{results}
Let us scale the exchange interaction components with the  unit of energy  $J$ as,
$$
r_n=\frac{J_n}{J}
$$ where $n=x,y,z$. Let us choose $r_z=1$, then $r_x,r_y$ can be used as the measure of the anisotropy in the exchange interaction. It can be seen from the definition of the function given in Eq. \re{denk7} that the transformation $J_x\rightarrow J_y,J_y\rightarrow J_x$ does not change the function. Hence, we can say that mentioned transformation does not affect the formulation. Because of that let us fix $r_x$  and concentrate only on varying $r_y$ values. Our investigation is on the simple cubic ($z_0=5,z_1=0$) and body centered cubic ($z_0=7,z_1=0$) lattices.

\subsection{Bimodal Distribution}

This distribution is given by Eq. \re{denk2} with $p=0$ and it distributes the longitudinal magnetic fields $\pm H_0$ to the lattice sites with equal percentages. Increasing randomness (which comes from increasing $H_0$ values in bimodal distribution) will reduce the critical temperature, as expected. Beside this, in order to concentrate on the variation of the  tricritical point  as a function of anisotropy in the exchange interaction, let us examine the evolution of the phase diagrams in a $(k_BT_c/J,H_0/J)$ plane with different $r_y$ values. In Fig. \re{sek1}, we can see the phase diagrams of the anisotropic quantum Heisenberg model on simple cubic  and body centered cubic lattices for a bimodal random magnetic field distribution in a $(k_BT_c/J,H_0/J)$ plane for some selected values of the $r_y$. We can track the path of the tricritical point (denoted by filled circles) with changing $r_y$ values for simple cubic and body centered cubic lattice in Figs. \re{sek1} (a) and \re{sek1} (b), respectively. Increasing $r_y$ values -which means that the anisotropy in the spin-spin interaction increases- reduce the two coordinates of the tricritical point in $(k_BT_c/J,H_0/J)$  plane. Also for arbitrarily fixed $H_0/J$ values, rising anisotropy reduces the critical temperature, as expected. For a certain $r_y$ value, we can see that tricritical point of the body centered lattice appears to be greater than that of the simple cubic lattice. We can also see from Fig. \re{sek1} that falling rate of a critical temperature corresponding to a certain $H_0/J$ value rises with increasing $r_y$. Tricritical point coordinates of the isotropic case ($r_x=r_y=1$) found in the present work $(H_0/J,k_BT_c/J)=(2.274,2.748)$ can be compared with the classical case $(H_0/J,k_BT_c/J)=(2.389,2.785)$ \cite{ref14,ref16}. Our values in the quantum case are slightly lower than corresponding classical values, as expected.

\begin{figure}[h]\begin{center}
\epsfig{file=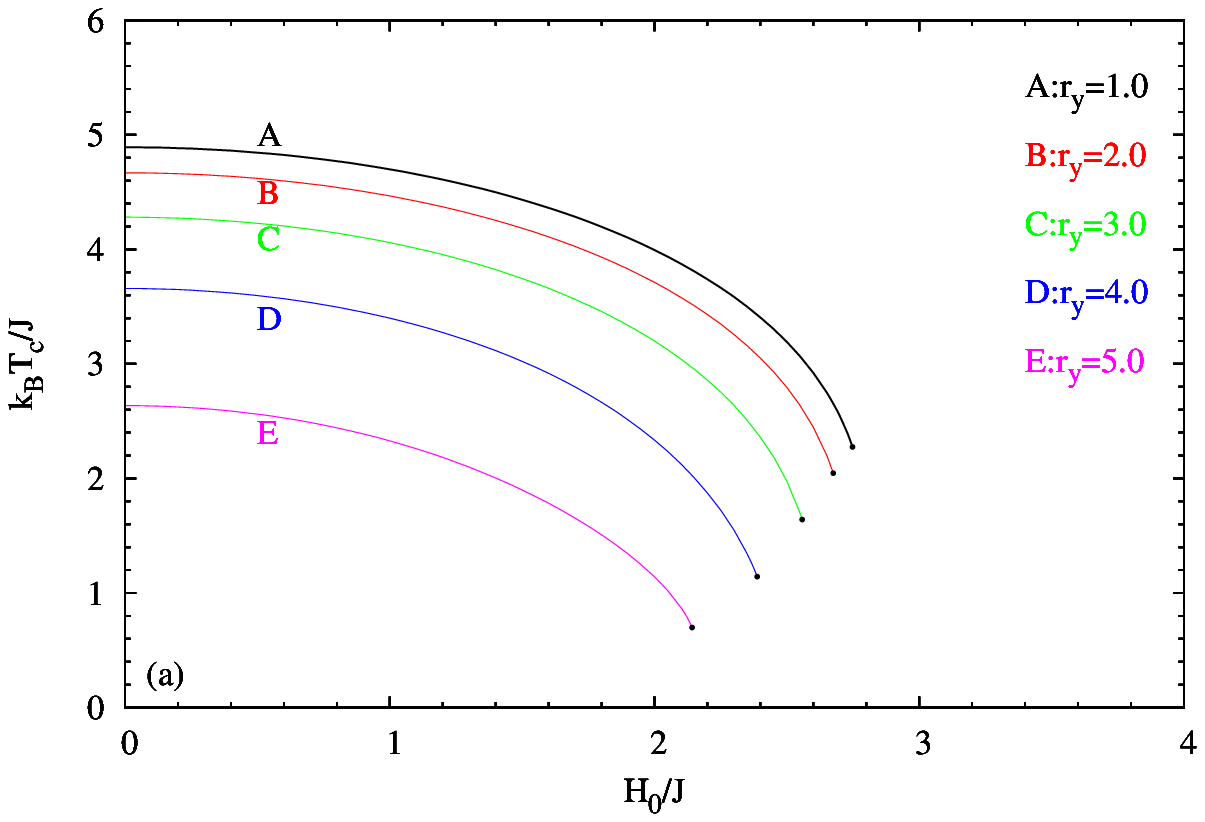, width=7.0cm}
\epsfig{file=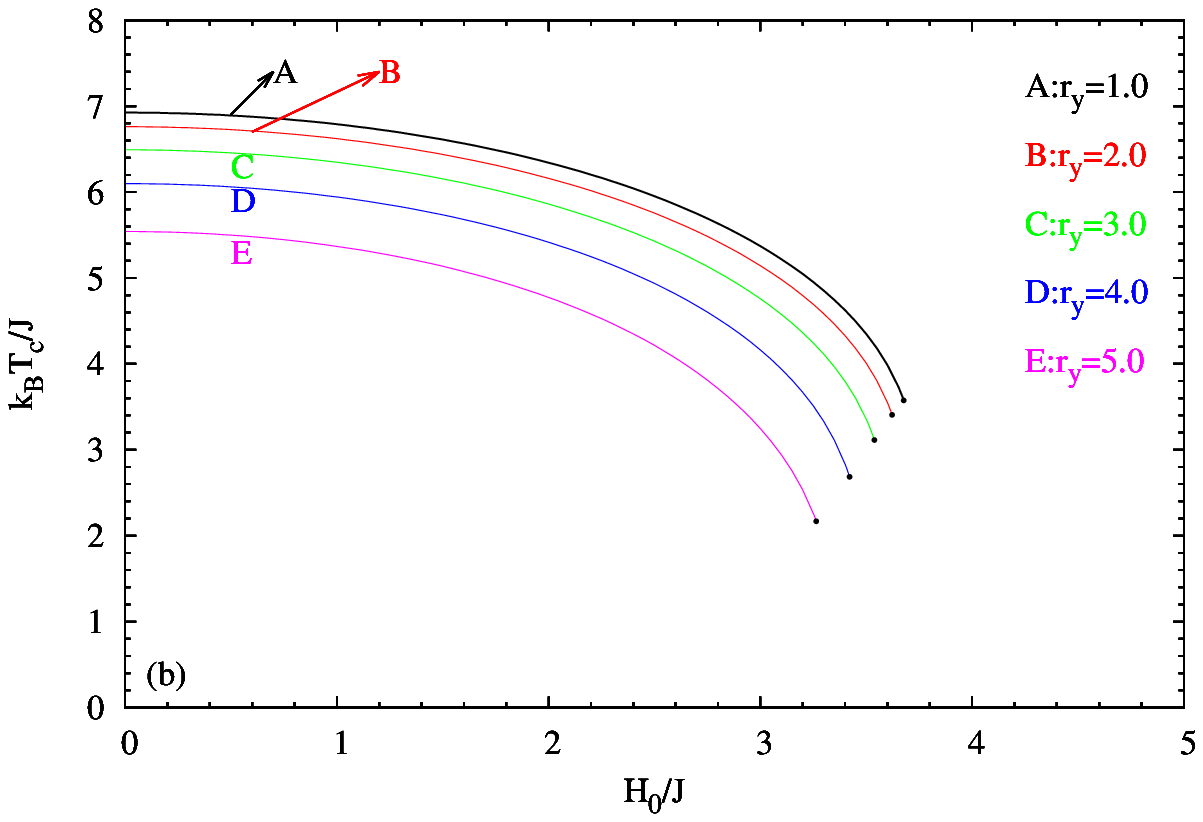, width=7.0cm}
\end{center}
\caption{Phase diagrams of the anisotropic quantum Heisenberg model for (a) simple cubic  and (b) body centered cubic lattices with the bimodal random magnetic field distribution in $(k_BT_c/J,H_0/J)$ plane for some selected values of $r_y$. Solid lines represent the second order transitions, while the filled circles denote the tricritical points. Fixed parameter value is $r_x=1.0$.} \label{sek1}\end{figure}

\subsection{Trimodal Distribution}

As we can see from Eq. \re{denk2} that trimodal distribution brings the system closer to the pure one (system with zero magnetic field) with increasing $p$. Thus, it is expected that the phase diagrams in  $(k_BT_c/J,H_0/J)$ plane with increasing $p$ will become parallel lines with the $H_0/J$ axis (at the value of $p=1$), which is nothing but just the critical temperature of the pure system with given parameter values. Also for a given $H_0/J$ value, increasing $p$ means decreasing randomness effect. Therefore, the ferromagnetic region is expected to get expanded in  $(k_BT_c/J,H_0/J)$  plane. However, it would be interesting to inspect again the evolution of the tricritical point with increasing $p$ values. In Fig. \re{sek2}, we can see the phase diagrams of the anisotropic quantum Heisenberg model in $(k_BT_c/J,H_0/J)$ plane for simple cubic  and  body centered cubic lattices in the presence of a trimodal random magnetic field distribution, and for some selected values of $p$. We can see from Fig. \re{sek2} that observed tricritical points at low $p$ values gradually decrease and reduce to $k_BT_c/J=0$ at a certain $H_0/J$. Right after the disappearance of the tricritical point, the system exhibits a second order reentrant behavior, i.e. after a certain value of $p$  (which depends on the lattice geometry, as well as $r_x,r_y$ values) the system which stays in a disordered phase at zero temperature can pass to an ordered phase with a second order transition due to increasing thermal fluctuations, then it passes again to another disordered phase characterized by a second order transition with increasing temperature. For $p$ values greater than a special value (let us denote it by $p^{*}$), the phase diagrams in $(k_BT_c/J,H_0/J)$ plane do not intersect the x-axis, i.e. for $p>p^*$, the system will stay in an ordered phase at zero temperature for any $H_0/J$ value. This specific value depends on the lattice geometry and the degree of the anisotropy in the exchange interaction  ($r_x$ and $r_y$). When we compare Figs. \re{sek2}(a) and (b) with each other, we can conclude that increasing exchange anisotropy does not alter this situation significantly for small anisotropy, but it affects only the critical values (e.g. compare Figs. \re{sek2} (a) and (b) or Figs. \re{sek2} (c) and (d)). The same observation also holds for a body centered cubic lattice. After all, we can say that in general, three qualitatively different regions are observed in the phase diagrams depicted in a $(k_BT_c/J,H_0/J)$ plane as $p$ rises. Namely, a tricritical behavior region, a second order reentrant region, as well as a region which gets expanded towards increasing $H_{0}/J$ direction. We also note that the second order reentrant region corresponding to the simple cubic lattice is wider than that of the body centered cubic lattice. This point is depicted more apparently in Fig. \re{sek3}.

In order to see the dependence of $p^*$ on the exchange anisotropy, we plot the variation of $p^*$ with $r_y$ for some selected values of $r_x$ in Figs. \re{sek3} (a) and (b) for simple cubic and body centered cubic lattices, respectively. We note that for a certain value of $r_x$, the points which lie below the  related curve in $(p^*,r_y)$ plane have disordered ground states for high $H_0/J$ values. In other words, for a selected $r_x$ value, if we choose $(p,r_y)$ pairs which lie under the curve corresponding to selected $r_x$, this means that the phase diagram of the system in $(k_BT_c/J,H_0/J)$ plane either exhibits a tricritical behavior or a second order reentrant behavior.  For a system with parameter values of ($p,r_y$) which lie above the related curve has an ordered ground state in high $H_0/J$ region, i.e. the phase diagram of the system in a $(k_BT_c/J,H_0/J)$ plane does not intersect the x-axis. Thus, we can see from Fig. \re{sek3} that increasing anisotropy contracts the region where the phase diagrams on the right side of the $(k_BT_c/J,H_0/J)$ plane are stretched. For a fixed value of $r_x$, and after a certain value of $r_y$, we get $p^*=1.0$ which means that the system can not exhibit an ordered ground state at high values of $H_0/J$ regardless the value of $p$. Another point is that the simple cubic lattice has a higher $p^*$ value for a fixed $r_x,r_y$, unless we have $p^*=1.0$.

Finally, in order to reveal the relation between the $p^*$ value and the anisotropy in the exchange interaction more clearly, we depict equally valued $p^*$ curves in $(r_x,r_y)$ plane in Fig. \re{sek4}. As we can see from Fig. \re{sek4} that these curves are symmetric about the $r_x=r_y$ line.  This is obvious, since $r_x\rightarrow r_y, r_y\rightarrow r_x$ transformation does not alter the equations of the system, as stated in the beginning of this section. In general, for a selected value of $r_x$, if we move on the increasing direction of $r_y$, we pass through increasing $p^*$ curves, successively.  But for instance, if we choose $r_x=2.0$ and continue in increasing direction of $r_y$ starting from $r_y=1.0$ (in Fig. \re{sek4} (a)), we intersect $p^*=0.54$ curve two times and the other curves once. Thus, the relation between anisotropy and $p^*$ is not linear; i.e. we can not say $p^*$ increases monotonically with increasing anisotropy.

\begin{figure}[h]\begin{center}
\epsfig{file=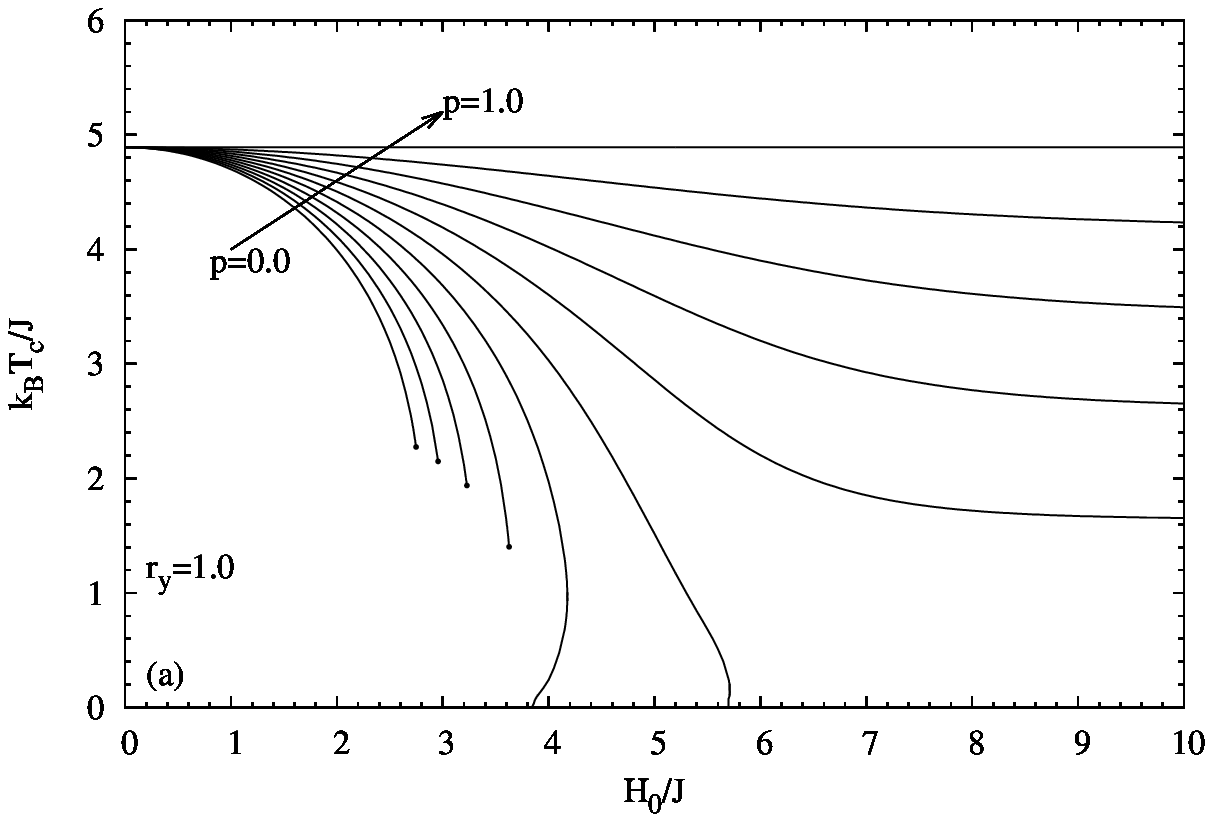, width=7.0cm}
\epsfig{file=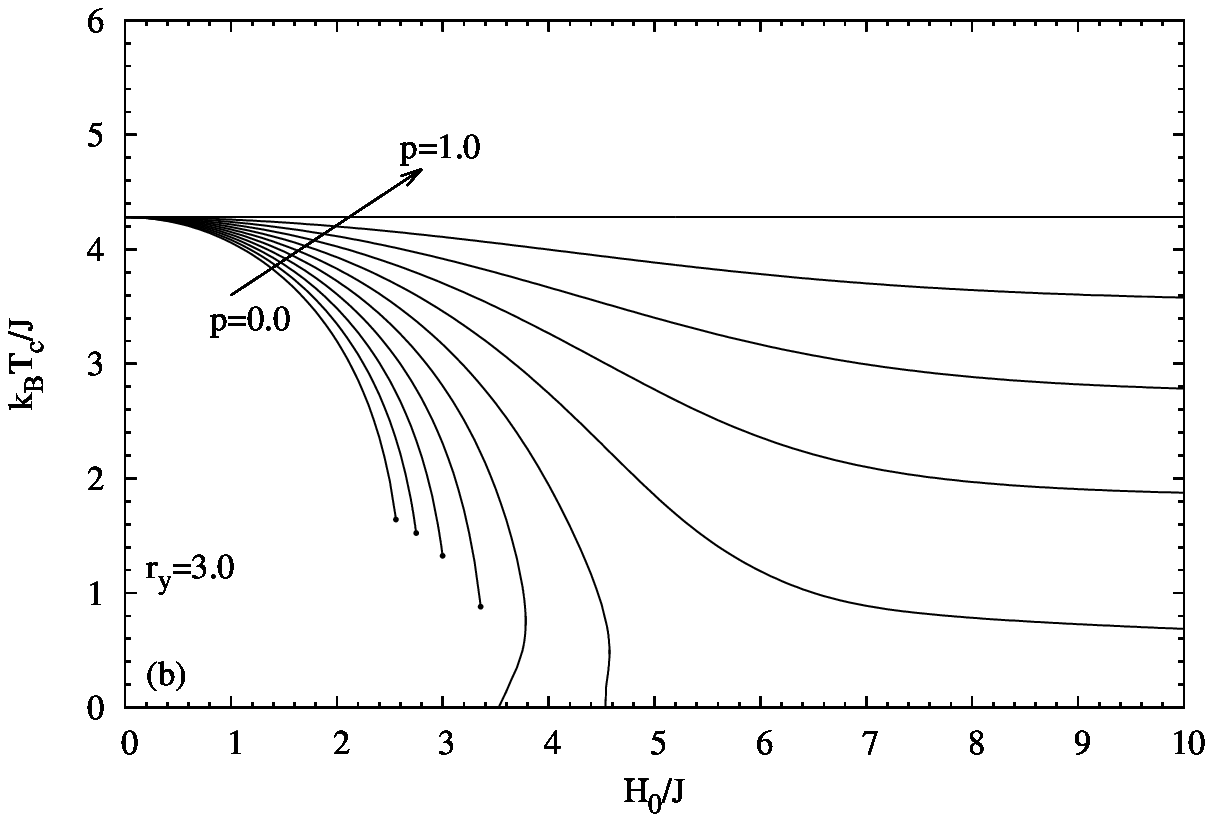, width=7.0cm}
\epsfig{file=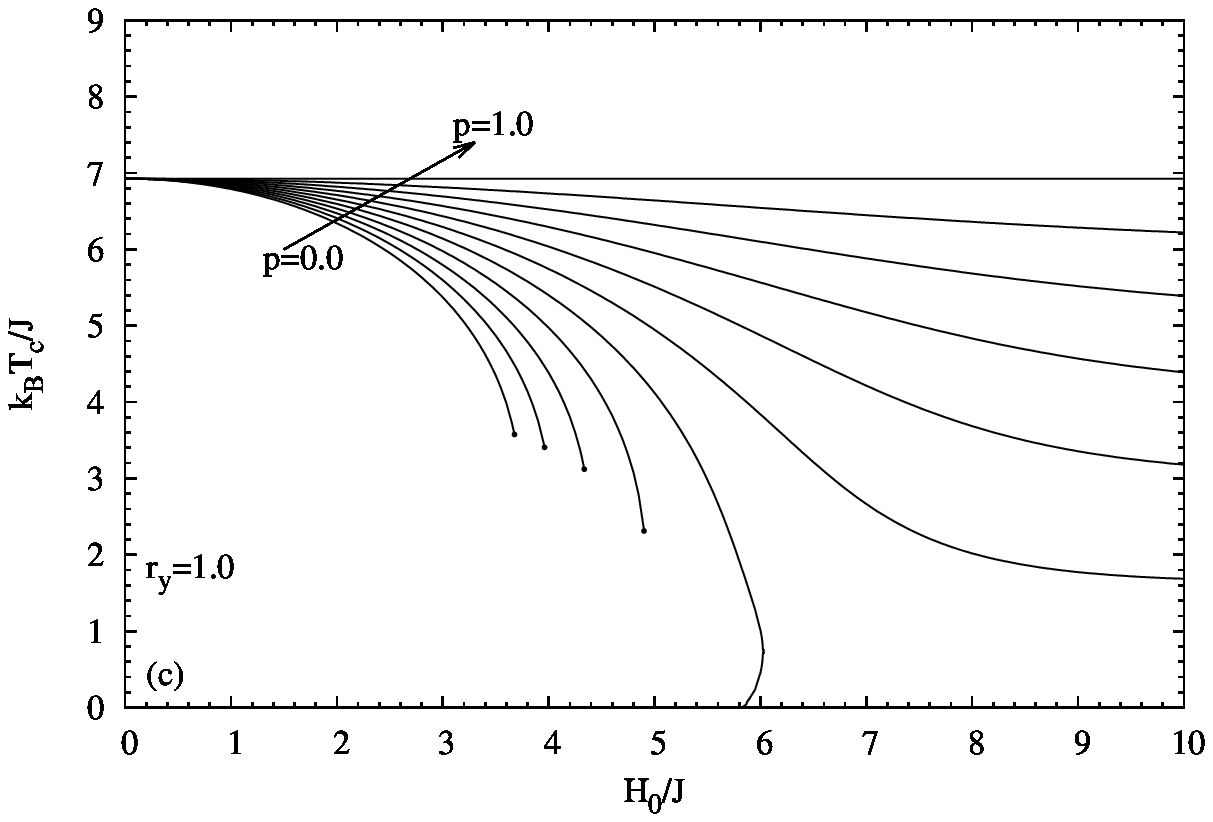, width=7.0cm}
\epsfig{file=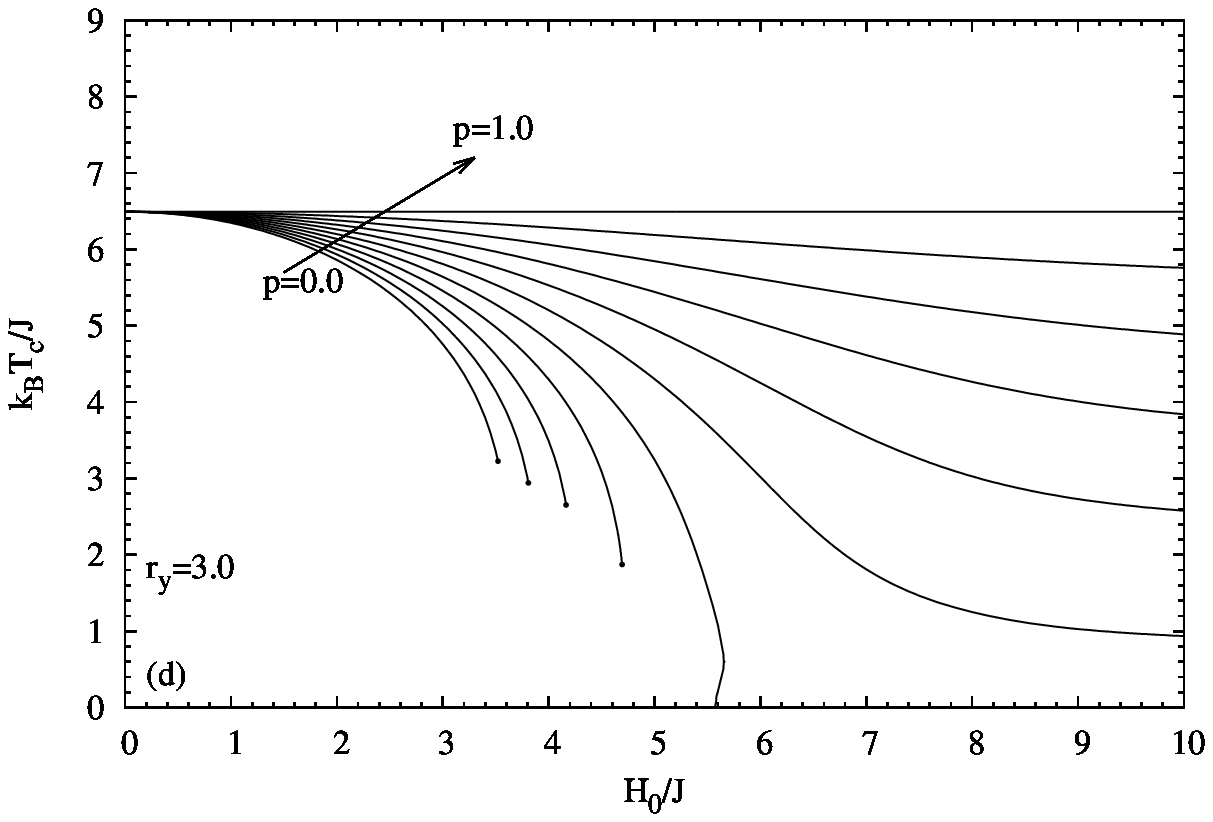, width=7.0cm}
\end{center}
\caption{Phase diagrams of the anisotropic quantum Heisenberg model for (a),(b) simple cubic  and (c),(d) body centered cubic lattices with a trimodal random magnetic field distribution in $(k_BT_c/J,H_0/J)$ plane for selected values of $p$ from $p=0.0$ to $p=1.0$ with increment of $0.1$. Increment direction is shown as arrows in each figure. Solid lines represent the second order transitions, filled circles denote the tricritical points. Fixed parameter value  is $r_x=1.0$.} \label{sek2}\end{figure}

\begin{figure}[h]\begin{center}
\epsfig{file=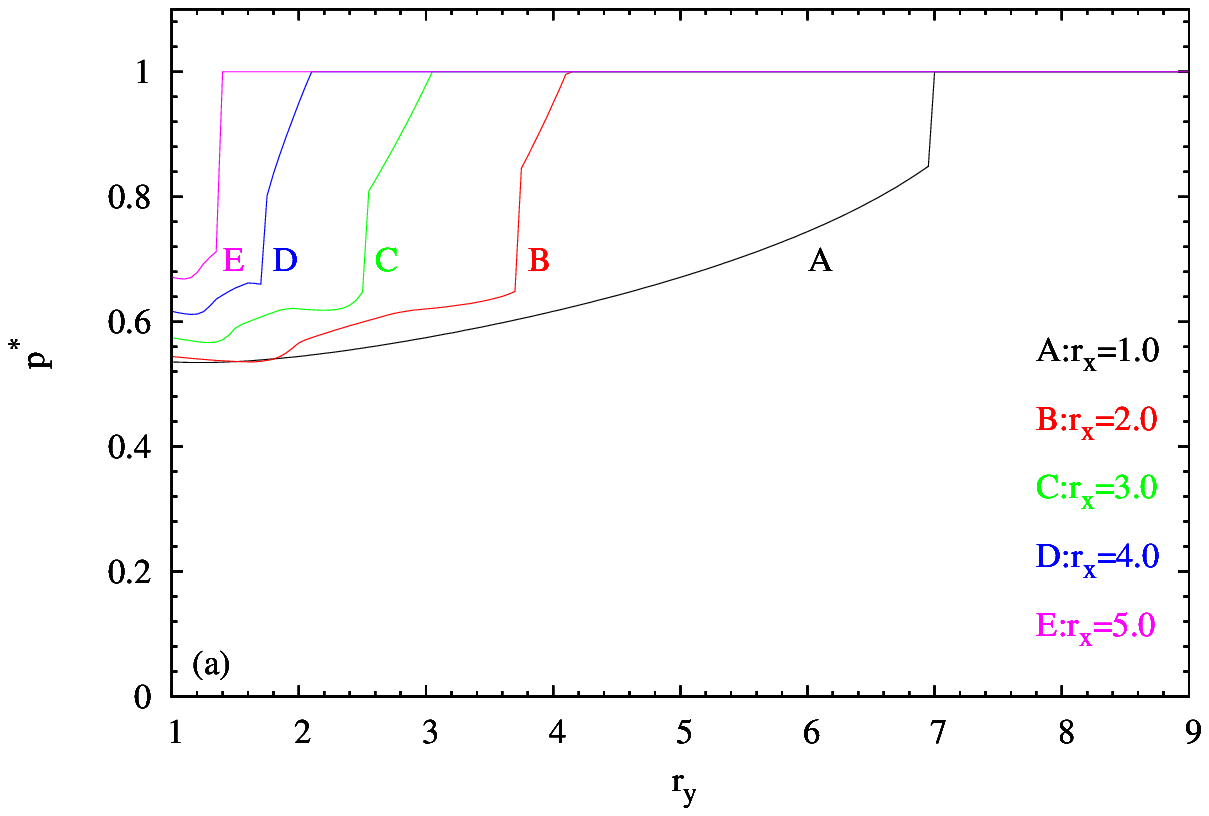, width=7.0cm}
\epsfig{file=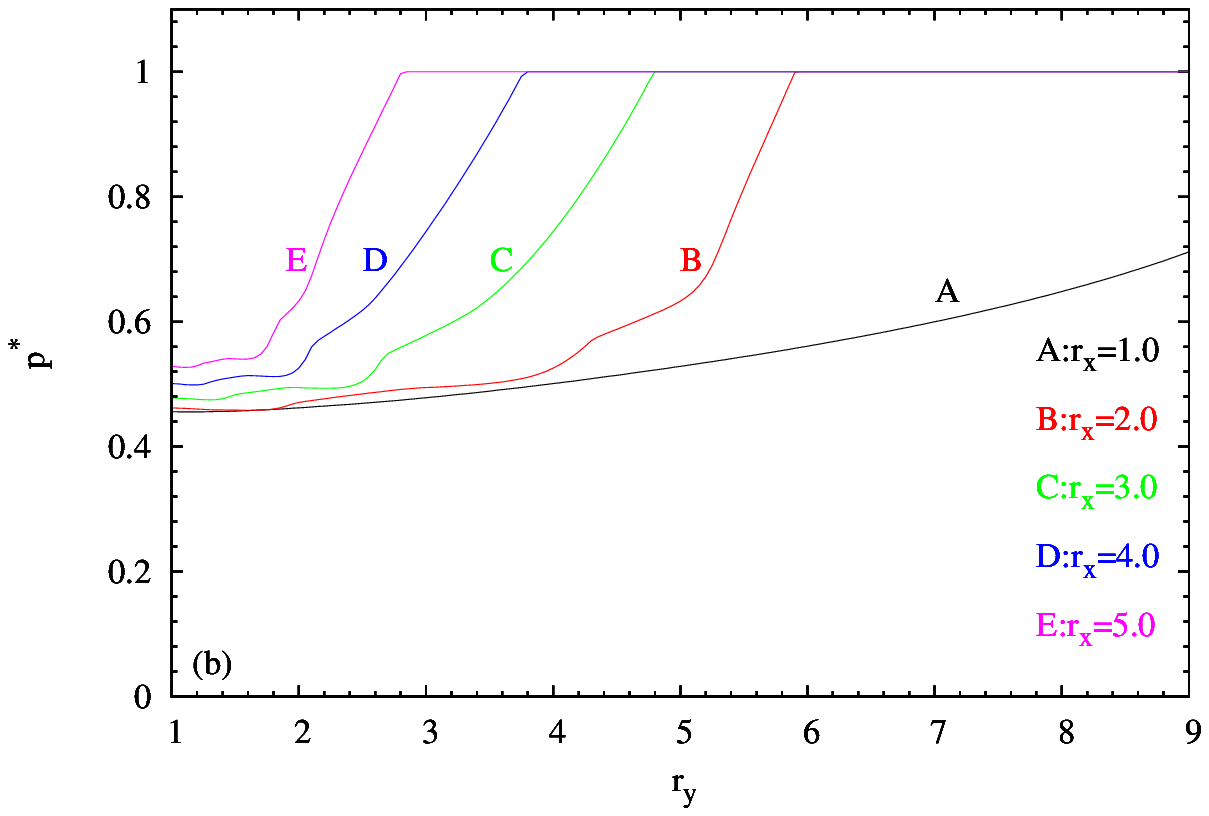, width=7.0cm}
\end{center}
\caption{Variation of $p^*$ with $r_y$ for some
selected values of $r_x$ for the anisotropic quantum Heisenberg model  with a trimodal random magnetic field for (a) simple cubic, and (b) body centered cubic lattice.} \label{sek3}\end{figure}

\begin{figure}[h]\begin{center}
\epsfig{file=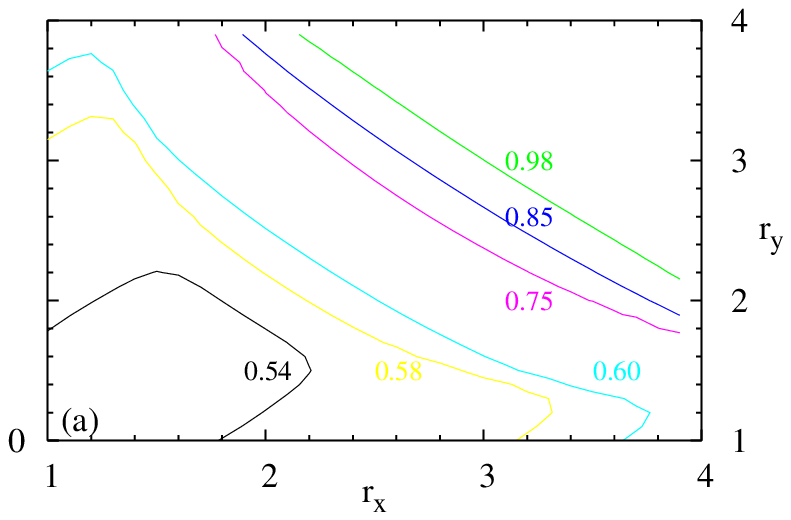, width=7.0cm}
\epsfig{file=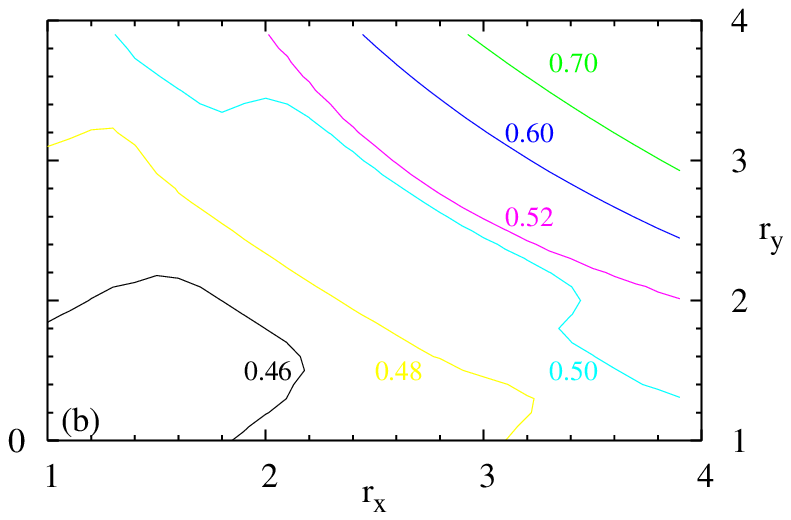, width=7.0cm}
\end{center}
\caption{Equally valued $p^*$ curves in  $(r_x,r_y)$ plane for the anisotropic quantum Heisenberg model  with a trimodal random magnetic field on (a) simple cubic, (b) body centered cubic lattice.} \label{sek4}\end{figure}

\section{Conclusion}\label{conclusion}

The effect of the trimodal random magnetic field distribution on the phase
diagrams of the anisotropic quantum Heisenberg model has been
investigated in detail. The effects of the random magnetic fields in the presence of exchange interaction anisotropy have been discussed for three dimensional lattices, namely for simple cubic and body centered cubic lattices. Qualitatively similar characteristics have been observed for the phase diagrams of these different lattice geometries. However, quantitative differences have been obtained for bimodal and trimodal distributions.

Calculations for bimodal distribution performed in this work may be thought as a generalization of the earlier works which were devoted to a bimodal distribution in isotropic Heisenberg model\cite{ref14,ref15,ref16,ref17} case to the anisotropic case. According to the phase diagrams plotted in $(k_BT_c/J,H_0/J)$ plane, presence of exchange anisotropy in the system reduces all critical temperatures, as well as tricritical points.

On the other hand, trimodal distribution gives rise to qualitatively three different phase diagram types with varying $p$ values  in $(k_BT_c/J,H_0/J)$ plane. Namely, as $p$ increases starting from $p=0.0$, we observe phase diagrams exhibiting tricritical behavior, phase diagrams with second order reentrant phenomenon, and  phase diagrams which become stretched to the right hand side of $(k_BT_c/J,H_0/J)$ plane, respectively. At the end of this evolution process (i.e. for $p=1.0$), phase diagrams become parallel lines with the $H_0/J$ axis. In this case, $k_BT_c/J$ value just corresponds to the critical temperature of the pure system. This classification scheme is valid for both simple cubic and  body centered cubic lattices.

Particular attention has been devoted on the $p$ value (namely $p^*$) for which the phase diagrams become stretched to the right hand side of $(k_BT_c/J,H_0/J)$ plane. The effect of the anisotropy in the exchange interaction on $p^*$ value has been investigated in detail. For this purpose, equally valued  $p^*$ curves have been obtained in $(r_x,r_y)$ plane. It has been shown that, the effect of the parameters $r_x$, $r_y$ on the value of $p^*$ is not linear.

We hope that the results  obtained in this work may be beneficial form both theoretical and experimental point of view.

\end{document}